\documentclass[11pt,aps,showpacs,showkeys ]{revtex4}
\usepackage{amsmath}
\usepackage[cmtip,arrow]{xy}
\usepackage{pb-diagram,pb-xy}
\usepackage{slashed}
\usepackage{amssymb,amsopn}
\def \be {\begin{equation}}
\def \ee {\end{equation}}
\def \bea {\begin{eqnarray}}
\def \eea {\end{eqnarray}}

\usepackage{graphicx}
\begin{document}
\title{Polymer quantization, stability and higher-order time derivative terms}
\author{ Patricio Cumsille$^{1,2}$}
\author{ Carlos M. Reyes$^{1}$}
\email[Electronic mail: ]{creyes@ubiobio.cl}
\author{Sebastian Ossandon$^{3}$ }
\author{ Camilo Reyes$^{4}$}
\affiliation{$^{1}$ Departamento de Ciencias B{\'a}sicas, Universidad 
del B{\'i}o B{\'i}o, Casilla 447, Chill\'an, Chile}
\affiliation{$^{2}$ Centro de Biotecnolog\'ia y Bioingenier\'ia (CeBiB), Universidad 
de Chile, Beaucheff 851, Santiago de Chile}
\affiliation{$^{3}$ Instituto de Matem\'aticas, Pontificia Universidad 
Cat\'olica de Valpara\'iso, Casilla 4059, Valpara\'iso, Chile}
\affiliation{$^{4}$ Departamento
de Ciencias Fisicas, Facultad de Ciencias Exactas, Universidad Andres Bello, Republica 220, 
Santiago, Chile}
\begin{abstract}
The possibility that fundamental discreteness implicit in a quantum gravity theory may
act as a natural regulator 
for ultraviolet singularities arising in quantum field theory has been intensively studied. 
Here, along the same expectations,
we investigate whether a nonstandard representation, called polymer
representation can smooth away 
the large amount of negative energy that 
afflicts the Hamiltonians of
higher-order time derivative theories; rendering the theory unstable when interactions come into play.
We focus on the fourth-order
 Pais-Uhlenbeck model which can be reexpressed as the sum of two 
 decoupled harmonic oscillators one producing positive energy and the other negative energy.
 As expected, the Schr\"odinger quantization of such model
  leads to the stability problem or to negative norm states called ghosts.
Within the framework of polymer quantization we show the existence of new regions where
the Hamiltonian can be defined well bounded from below.
 \end{abstract}
 \pacs{03.65.-w, 04.60.Ds, 04.60.Nc, 04.60.Pp }
 \keywords{higher time derivatives; polymer quantization; stability}
\maketitle
\section{Introduction}
The standard model of particles has foundations on local quantum field theories 
having operators of mass dimension $d\leq4$. 
These operators are
justified in order to implement
  the requirements of 
 stability and unitarity without further elaborations~\cite{eliezer}.
However, when going to 
 higher energies it is commonly believed 
 that
higher-order operators will play a key role in
describing fundamental physics. 
This may be particularly true when they involve higher time derivatives
since the new modes
that arise allow
to describe effects from a high scale.
Usually these new modes are very high
when the coupling of the higher-order time derivatives are suppressed by the high scale
as taken in effective approaches.
Higher-order operators have received increased attention over the years. They
have been investigated in the context of
 loop quantum gravity~\cite{Gambini,Alfaro,Alfaro1,Alfaro2,Sahlmann,Sahlmann1},
Lorentz symmetry violation~\cite{KM1,KM2,KM3,MP,schreck,schreck1}, causality 
and stability \cite{reyescausality}, 
fine-tuning~\cite{large,Reyes,Reyes1,Reyes2}, the hierarchy problem~\cite{Grinstein,Grinstein1}, 
radiative corrections~\cite{radcorr,radcorr1,radcorr2} and nonminimal 
couplings~\cite{casana-manoel,casana-manoel1,casana-manoel2,casana-manoel3,petrov,petrov1}. The presence of
higher-order derivatives 
in the gravitational sector are a key ingredient in order to achieve a consistent 
renormalization in the
semiclassical approach, where matter fields are quantized over classical curved background~\cite{HD-grav}.

Higher time derivative theories were introduced long time ago by 
Ostrogradsky~\cite{OSTRO}. The approach is based on 
a variational formalism and involves
a  Lagrangian $L(x,\dot x, \ddot x \dots,  
 x^{(n)}=\frac{d^nx}{dt^n})$ and an
 extended Hamiltonian $H(x,\pi_{\dot x}, \pi_{\ddot x }\dots,  
 \pi_{x^{(n)}})$ of $2n$
variables. 
Ostrogradsky showed 
that the formalism leads a higher-order Hamiltonian 
 not bounded from below as can be seen in the 
   non-quadratic momenta terms that appear.
This is the classical Ostrogradsky instability of higher-order time derivative theories
which can be avoided in a few cases, for instance when the 
models has constraints~\cite{mijael}.

The quantization of higher time derivative theories can be implemented 
introducing a change of variables in order to transform the 
original Lagrangian into a sum of decoupled normal-order Lagrangians. In general
one of these Lagrangians has large negative energy
leading to the instability or alternatively, by changing the vacuum state, 
 to an indefinite metric theory~\cite{P-U}.
The instability of the Hamiltonian has received much attention and has been tackled
 from 
 different perspectives, such as phase space reduction~\cite{cheng,cheng1,pert}, 
 complex canonical transformations~\cite{hugo}, $\mathcal {PT}$ 
 symmetry~\cite{PT,PT1,PT2}, and Euclidean-path reduced amplitudes~\cite{hawkings}, gravitational ghosts and 
 tachyons~\cite{grav-ghosts,grav-ghosts1}.
In quantum field theory,  Lee and Wick showed that by imposing
 the negative norm states to decay
 it is possible 
 to preserve unitarity order by order in perturbation theory~\cite{L-W,L-W1}.
Resurgence of such ideas have been used 
to solve the hierarchy problem~\cite{Grinstein,Grinstein1} and 
in higher-order effective models with Lorentz symmetry violation~\cite{unit,unit1,unit2}.

In this work we study the stability of higher-order time derivative
models within the framework of polymer quantization. In particular,
we focus on the Pais-Uhlenbeck (P-U) model.
The polymer representation is a non-standard representation of
 quantum mechanics
inspired by some results that emerge from loop
quantum gravity. The 
 possibility of space discreteness that appears in loop quantum gravity~\cite{ash}
  has served to improve the convergence of quantum field theories~\cite{Sahlmann,Sahlmann1}
 and cosmological singularities~\cite{Bojowald}. 
 In this paper our main goal is to test
whether the fundamental discreteness implicit
in the polymer representation allows to improve the stability of higher-order theories.
The polymer quantization has been considered in several 
studies such as two-point functions~\cite{propagators,propagators1,propagators2}, 
cosmology~\cite{Bojowald,tomasz}, central forces~\cite{r,r1}, 
higher space derivatives~\cite{martin}, thermodynamics~\cite{Chacon},
compact stars~\cite{Chacon-Acosta}
 and low energy limits~\cite{corichi,corichi1}.

The organization of this work is as follows. In section \ref{sectionII}
we introduce the P-U model and we explicitly show 
the origin of the instability in
 the Schr\"odinger 
quantization.
In the third section we give a basic review of 
the polymer formalism. In section \ref{sectionIV}
we polymer quantize the P-U model
and solve the Hamiltonian eigenvalue equation. From the previous results 
we analyze stability in the region of validity of the theory. In the last section we give the conclusions.
\section{The Pais-Uhlenbeck model}\label{sectionII}
The P-U model consists of an harmonic 
oscillator coupled to a higher-order term described by the Lagrangian
\begin{eqnarray}\label{Lagrangian}
 L=\frac{1}{2}m\dot x^2-\frac{1}{2}kx^2-\frac{g}{2}\ddot x^2\;,
\end{eqnarray}
where $g$ is a small parameter.  
The equation of motion is
\begin{equation}
\label{Eqm.P-U}
-kx-m \frac{d^2 x}{dt^2}-g \frac{d^4 x}{dt^4}=0\;.
\end{equation}
Inserting the plane wave ansatz $x(t)=x_0e^{-iy t}$ produces
the four solutions 
\begin{eqnarray}\label{omegas}
y=\pm \sqrt{ \frac{m\pm \sqrt{m^2-4kg}}{2g}}\;.
\end{eqnarray}
We define the two positive solutions according to
\begin{eqnarray}
 \omega_1 =\omega_0 \sqrt{ \frac{1- \sqrt{1-4\varepsilon}}{2\varepsilon}} \;,
\quad  \omega_2=\omega_0 \sqrt{ \frac{1+ \sqrt{1-4\varepsilon}}{2\varepsilon}} \;,
\end{eqnarray}
where we introduce $\omega_0=\sqrt{k/m}$ as the usual frequency and 
$g=\frac{m\varepsilon}{\omega_0^2}$ with $\varepsilon$
a small dimensionless parameter.
 Now, in the limit $\varepsilon \to 0$ 
 the first solution tends to the
usual harmonic solution 
$\omega_1 \to \omega_0 $ and the second one
$\omega_2\to  \frac{\omega_0}{ \sqrt{\varepsilon} }$ blows up. 
One expects this behavior of the last solution
 in theories with higher-order
 time derivative theories indicating a possible window to physics 
 at higher energy scales.
 In order to avoid imaginary solutions
 we impose 
$\varepsilon \leq \varepsilon_c =1/4$ where $\varepsilon_c$ is the critical value 
at which the two solutions collapse 
$\omega_1=\omega_2$.

The conjugate momenta to $x$ and 
 $\dot x$ are defined
  by the expressions
\begin{eqnarray} \label{mom2}
p&=&\frac{\partial L}{\partial \dot x}-\frac{d}{dt}\left
(\frac{\partial L}{\partial \ddot x}\right)\;,
\\ 
\pi&=&\frac{\partial L}{\partial \ddot x}\;.
\end{eqnarray}
From \eqref{Lagrangian}, they are given by
\begin{eqnarray} \label{mom2}
p&=&m \dot x+g  x^{(3)}\;, \\ \pi&=&-g \ddot x\;.
\end{eqnarray}
The Hamiltonian is
constructed via the extended Legendre transformation
$H=p\dot x+\pi \ddot x-L$ which after substitution
yields  
\begin{eqnarray}\label{H}
H=-\frac{\pi^2}{2g}  - \frac{m \dot x^2}{2}+\dot x p+\frac{k x^2}{2}\;.
\end{eqnarray}
Let us consider the new set of variables
\begin{eqnarray}\label{equis}
 x_{1}= \frac{\omega_2^2x-\pi/g}{\omega_2^2-\omega_1^2}\;, 
\qquad   x_{2}=-\frac{\omega_1^2x-\pi/g}{\omega_2^2-\omega_1^2}    \;,
\end{eqnarray}
and
\begin{eqnarray}\label{pes}
    p_{1}=  p-(m-g\omega_2^2) {\dot
 x }    \;, \qquad  p_{2}= p-(m-g\omega_1^2) {\dot x }\;,
\end{eqnarray}
where $ m^{\prime} = \sqrt{m^2-4kg}$. 
They allow to define the ladder variables 
\begin{eqnarray}\label{ladder}
 a_1 &=&  \sqrt{\frac{m^{\prime}\omega_1}{2\hbar}} \left(
 x_1+i {\frac{1}{ m^{\prime}\omega_1}}   p_1\right)\;, \nonumber \\
 a_2 &=&   \sqrt{\frac{m^{\prime}\omega_2}{2\hbar}}  \left(   x_2+i 
{\frac{1}{ m^{\prime}\omega_2}}    p_2 \right) \;.
\end{eqnarray}
Using these variables the Hamiltonian 
can be expressed as the sum of two decoupled harmonic oscillators
\begin{eqnarray}
 H=\frac{\hbar \omega_1}{2}(  a_1  a^{\dag}_1 +  a^{\dag}_1  a_1)
-\frac{\hbar \omega_2}{2}( a_2  a^{\dag}_2 +  a^{\dag}_2   a_2)\;,
\end{eqnarray}
where  we refer to the first and the second term as 
the positive and negative sectors of the theory.

The quantization of such model follows by imposing the usual commutation relations
 in the extended phase space 
$\left[ \hat x, \hat p\right]=i \hbar$ and  $[ \hat { \dot x}, \hat {\pi} ]=i \hbar$.
With the use of the canonical variables
$ \left[\, \hat x_{1},\, \hat p_{1}\,\right]=i \hbar$ and $ \left[ \, \hat x_{2},\, \hat p_{2}\,\right]=i\hbar $
one can check that 
the creation and annihilation operators satisfy
\begin{eqnarray} \label{ghost}
\left[ \hat a_1,\hat a_1^{\dag}\right]=1, \qquad \left[ \hat a_2,\hat a_2^{\dag}\right]=1 \;.
\end{eqnarray}
To find the ground state wave function denoted by $\Psi_0$, consider the 
 explicit action of the operators in the Hilbert space
\begin{eqnarray}
\hat x \psi=x \psi\;, \qquad \hat p \psi=-i\hbar\frac{ \partial \psi }{\partial x}\;,
\\
\hat {\dot x}\psi =\dot x \psi\;, \qquad \hat \pi \psi =-i\hbar\frac{ \partial \psi}{\partial {\dot x}}\;.
\end{eqnarray}
Inserting these expression into
 the previous creation and annihilation operators \eqref{ladder}
 we identify two realizations for the vacuum. The first one is to define
the vacuum as the one annihilated by $\hat a_{1}$
and $\hat a_{2}$
\begin{eqnarray}
\hat a_{1}\Psi_0=\hat a_{2}\Psi_0=0,
\end{eqnarray}
leading to 
\begin{eqnarray}\label{vacuum1}
\Psi_0= \mathcal N_0 \, e^{\frac{g(\omega_1-\omega_2)}{2\hbar  }(\dot x^2 +\omega_1
\omega_2 x^2)+\frac{ig}{\hbar} \omega_1 \omega_2x\dot x}\;,
\end{eqnarray}
where $\mathcal N_0$ is a normalization factor.
The Hamiltonian turns out to be 
\begin{eqnarray}
\hat H=\hbar \omega_1(\hat N_1+\frac{1}{2})
-\hbar \omega_2(\hat N_2+\frac{1}{2})\;,
\end{eqnarray}
where  $\hat N_1=\hat a^{\dag}_{1} \hat a_{1}$  and 
$\hat N_2=\hat a^{\dag}_{2} \hat a_{2}$ are the number operators of positive
and negative particles respectively.
We see that the energy is not bounded from 
below since one can always create more negative particles. 

A different vacuum $\Psi_0^{\prime}$ amounts to
change the annihilation operator in the negative sector and to maintain
the previous in the positive sector, namely
\begin{eqnarray}
\hat a_{1}\Psi^{\prime}_0=\hat a_{2}^{\prime}\Psi^{\prime}_0=0\;,
\end{eqnarray}
with $\hat a_{2}^{\prime}=\hat a_{2}^{\dag}$.
The vacuum state $\Psi_0^{\prime}$ in this case is
\begin{eqnarray}
\Psi_0^{\prime}= \mathcal N_0 \, e^{\frac{g(\omega_1+\omega_2)}{2\hbar  }(\dot x^2 -\omega_1
\omega_2 x^2)-\frac{ig}{\hbar} \omega_1\omega_2x\dot x}\;.
\end{eqnarray}
Note that $\Psi_0^{\prime}$ can be obtained in \eqref{vacuum1} performing the transformation 
 $\omega_2 \to  -\omega_2$.
The Hamiltonian is found to be
\begin{eqnarray}
 \hat H^{\prime}=\hbar \omega_1( \hat N_1+\frac{1}{2})
+\hbar \omega_2(\hat N_2^{\prime}+\frac{1}{2})\;,
\end{eqnarray}
where now $[ \hat a^{\prime}_2,\hat a_2^{\prime \dag}]=-1$ 
and $\hat N_2^{\prime}=  -  \hat a^{\prime \dag}_{2} \hat a^{\prime}_{2}$ is 
the new number operator.
In this case the theory has positive defined 
Hamiltonian but the price to pay is to 
end up with
 negative norm states or ghosts
that may threaten the conservation of unitarity
and with non-normalizable wave functions.
\section{Polymer representation}\label{sectionIII}
In quantum mechanics when dealing with the adjoint
operators $\hat x$ and $\hat p$ 
usually one encounters some technical problems due to their unboundedness.
Therefore, it is convenient to switch to 
the exponentiated versions $\hat U(\alpha)$
and $\hat V(\beta)$ 
\begin{eqnarray}\label{def-op}
\hat U(\alpha)=e^{i\alpha \hat x}, \qquad \hat V(\beta)=e^{i\beta \hat p/ \hbar}\;,
\end{eqnarray}
whose action are defined by
\begin{eqnarray}\label{UV}
\hat U(\alpha)  \psi (x)=e^{i\alpha x} \psi (x)\;, \qquad \hat V(\beta) \psi (x)= \psi (x+\beta)\;,
\end{eqnarray}
for all state $\psi(x)$ in the Hilbert space $ L^2({\mathbb R})$. Both operators $\hat U(\alpha)$ and $\hat V(\beta)$ satisfy
the Weyl-Heisenberg algebra
\begin{eqnarray}
\hat U(\alpha)  \hat U(\alpha^{\prime}) &=& \hat U(\alpha+\alpha^{\prime})\;, \nonumber\\
\hat V(\beta) \hat V(\beta^{\prime}) &=& \hat V(\beta+\beta^{\prime})\;, \nonumber\\
\hat U(\alpha)  \hat V(\beta) &=& e^{-i\alpha \beta}  \hat V(\beta)  \hat U(\alpha)\;, \label{algebra}
\end{eqnarray}
where the parameters $\alpha$ and $\beta$ 
have $(\text{length})^{-1}$  and length dimensions respectively. 
From the above algebra one can obtain the usual commutations relations
$[\hat x,\hat p]=i\hbar$.
Due to the Stone-von-Neumann theorem any
representation of the commutation relations have the form of the operators \eqref{def-op}, 
modulo unitarity transformation,
 since the two
operators $\hat U(\alpha)$
and $\hat V(\beta)$ are strongly continuous in their parameters, see~\cite{ash,Wald} and references therein.

In the polymeric construction one starts with  
a graph given by a countable set of points in the real line, denoted by 
$\gamma=\{x_j: j\in \mathbb N\}$, with some requirements~\cite{ash}. 
We define the functions associated to a graph
$\gamma$ as
\begin{eqnarray} 
f(x)= \left \{ 
\begin{array} { ll  } f_j & x=x_j \\
0 & x\not \in \gamma
\end{array} \right.   
\end{eqnarray}
and their Fourier transform functions $f(k)$ given by
\begin{eqnarray}
f(k)=\sum_j f_je^{-ix_jk}\;,
\end{eqnarray}
 satisfying the relation
\begin{eqnarray}
\sum_j |f_j |^2 x_j^{2n}<\infty \; \text{for } n=0,1,2,\dots
\end{eqnarray}
We denote by $Cyl_{\gamma}$ the space of all cylindrical functions $f(k)$
and
$Cyl$  
the union of all $Cyl_{\gamma}$ over all graphs $\gamma$.
We add to the space $Cyl$ all the limits of 
Cauchy sequences, that is the Cauchy completion which is called
the polymeric Hilbert space denoted by $H_{poly}$ endowed with
 the scalar product 
\begin{equation}\label{Kronec}
\langle e^{-ikx_i} | e^{-ikx_j}  \rangle=\delta_{x_i,x_j}.
\end{equation}
Recall, this is an alternative form to view the Hilbert space 
within the construction of the rigged Hilbert space
 $\Omega \subset H \subset  \Omega^*$, see Ref.~\cite{ballentine}.

The main differences between the
 polymeric representation and the usual of quantum mechanics
 is that $H_{poly}$ is a non-separable space and 
 has an intrinsic fundamental discreteness
leading to a nonequivalent representation, see Ref.~\cite{separability}. 
To be precise, the action of the operators
$\hat U(\alpha)$ and $\hat V(\beta)$ 
given in Eq. \eqref{UV}  is
well-defined  in $H_{poly}$, however, in the polymer representation there is no
  self-adjoint 
 operator $\hat p=-i\hbar \frac{ \partial}{\partial x}$ such 
 that the second equality in \eqref{def-op} 
 is satisfied, that is to say, the momentum operator $\hat p$ 
 is not well defined on $H_{poly}$.
 This is due to the fact that $\hat V(\beta)$ is not weakly 
 continuous in the parameter $\beta$, as can be verified 
 using the modified product 
 with the Kronecker delta in Eq. \eqref{Kronec}.
Nevertheless,  
one can approximate the operator $\hat p$ with the expression 
\begin{eqnarray}\label{mom-pol} 
\hat p = -\frac{i\hbar}{\mu_0} \left( \hat V(\mu_0 /2)-\hat V(-\mu_0/2) \right) \;,
\end{eqnarray}
where $\mu_0$ is a fundamental length scale associated 
with a possible discreteness of space, coming from a 
more fundamental theory.
The above approximation is natural, at least in the 
distributional sense, since if we take the limit as $\mu_0 \to 0$ 
we recover the usual momentum operator in $L^2(\mathbb R)$.
\section{Stability and higher-order time derivatives}\label{sectionIV}
In section \ref{sectionII}  we have expressed 
the P-U Hamiltonian as two decoupled harmonic oscillators, for instance $H=H_1-H_2$.
Their quantum counterparts represent normal particles and nonstandard ones producing negative energy
sometimes called Lee-Wick particles~\cite{L-W,L-W1}. 
Using the new set of variables \eqref{equis} and \eqref{pes} we find 
\begin{eqnarray}
H_1&=&\frac{1}{2}k_1 x_1^2+\frac{1}{2m^{\prime}}p_1^2\;,\nonumber 
\\
H_2&=&\frac{1}{2}k_2 x_2^2+\frac{1}{2m^{\prime}}p_2^2\;,
\end{eqnarray}
where $k_j = m^{\prime} \omega_j^2$ with $j=1,2$. 
In other words, 
the P-U model
involves two oscillators with the same mass $m^{\prime}$ 
and so taking advantage of this fact we polymer quantize
the system as two individual harmonic oscillators.

The polymer Hilbert space $H_{poly}=H_{poly}(x_1) \otimes H_{poly}(x_2)$
comprises the polymeric spaces 
 for each oscillator. In the Hilbert space
we have 
the action of the operators 
\begin{eqnarray}
\hat U_1(\alpha_1)  \psi (x_1)&=&e^{i\alpha_1 x_1} \psi (x_1)\;,  \nonumber 
\\ \hat V_1(\beta_1)  \psi (x_1)&=& \psi (x_1+\beta_1)\;, \nonumber 
\\
\hat U_2(\alpha_2)  \psi (x_2)&=&e^{i\alpha_2 x_2} \psi (x_2)\;, \nonumber 
\\ \hat V_2(\beta_2)  \psi (x_2)&=& \psi (x_2+\beta_2)\;.
\end{eqnarray}
Considering the wave function $\psi(x_1,x_2) 
\in H_{poly}$ we arrive at the Schr\"odinger equation
\begin{eqnarray}
&&\left(\frac{1}{2}k_1 \hat x_1^2+\frac{1}{2m^{\prime}} \hat p_1^2- \frac{1}{2}k_2
 \hat x_2^2-\frac{1}{2m^{\prime}}\hat p_2^2 \right)\psi(x_1,x_2)\nonumber  \\&&=E \psi(x_1,x_2)   \;,
\end{eqnarray}
where $E$ is the total energy of the system and
 the momentum operators are
\begin{eqnarray}\label{p1}
\hat p_1 &=& -\frac{i\hbar}{\mu_1} \left( \hat V_1(\mu_1 /2)-\hat V_1 (-\mu_1/2) \right) \;,
\nonumber \\
\hat p_{2}& =& -\frac{i\hbar}{\mu_2} \left( \hat V_2 (\mu_2 /2)-\hat V_2(-\mu_2/2) \right) \;,
\end{eqnarray}
with $\mu_1$ and $\mu_2$ the fundamental lengths associated to 
$H_{poly}(x_1)$ and $H_{poly}(x_2)$. 

With the ansatz $\psi(x_1,x_2)=\psi_1(x_1) \psi_2(x_2)$
 and considering the cylindrical function for each oscillator $j$, namely
\begin{eqnarray}
\psi_j(k)=\sum_{\ell} \psi_j(x_{j,{\ell}}) e^{-ix_{j,{\ell}}k} \;, 
\end{eqnarray}
together with
 Eqs \eqref{p1}, we obtain 
the equations for the coefficients
\begin{eqnarray}
&&\psi_j(x_{j,{\ell}}+\mu_j)+\psi_j(x_{j,{\ell}}-\mu_j) \nonumber \\&&= 
\left(  2-\frac{2E_j\mu_j^2}{\hbar \omega_j d_j^2}
 + \frac{ \mu_j^2 x_{j,{\ell}}^2 }{ d_j^4 }\right) \psi_j( x_{j,{\ell}} )\;, \label{psi1} 
\end{eqnarray}
where we have introduced $E=E_1-E_2$
and $d_j^2 = \frac{\hbar}{m^{\prime} \omega _j}$.
The previous equations suggest that one can find solutions 
supported at the uniformly spaced points 
$x_{j,{\ell}} = x_0 + {\ell} \mu_j$
for some $x_0 \in [0,\min\{ \mu_j \} )$. Indeed, 
given the parameters of the equations 
$\{ \psi ( x_{j,{\ell}} ) \}_{{\ell} = -\infty}^{\infty}$, by 
using \eqref{psi1} one can construct a 
solution $\psi_j$ supported on the lattice 
$\gamma^{j}_{x_0,\mu_j}:=\{ x_{j,\ell} \in \mathbb R \: | \: x_{j,\ell}=
x_0 + \ell \mu_j, \: \ell \in \mathbb Z \}$.
Using this in Eq. \eqref{psi1} yields
\begin{eqnarray}\label{eq1}
 &&\psi_j( x_{j,{\ell}+1} ) + \psi_j( x_{j,{\ell}-1} )\nonumber  \\ &&= \left(  2 - 
\frac{2E_j \mu_j^2}{\hbar \omega_j d_j^2}  + \frac{\mu_j^2 \left( x_0 + {\ell} \mu_j 
 \right)^2 }{d_j^4} \right) \psi_j( x_{j,{\ell}} ) \;. 
\end{eqnarray}
Without loss of generality, let us consider both graphs based on the point 
$x_0=0$. Hence, with $x_{j,{\ell}}= {\ell} \mu_j$,
$\psi_j(k_j) = \sum_{{\ell}=-\infty}^{\infty} \psi_j( x_{j,{\ell}} ) e^{-ik_j {\ell} \mu_j}$ and
multiplying the equation \eqref{eq1} by $e^{-i k {\ell} \mu_j}$ and summing over ${\ell}$, we arrive at
\begin{eqnarray}
2\cos(k_j \mu_j)  \psi_j(k_j) = 2 \left( 1 - \frac {E_j \mu_j^2}{\hbar 
\omega_j d_j^2} \right) \psi_j(k_j) -\frac {\mu_j^2}{d_j^4} \psi_j''(k_j)\;,
\end{eqnarray}
where $k_j \in (-\frac{\pi}{\mu_j},\frac{\pi}{\mu_j} ) $.
Normalizing and arranging the terms we get
\begin{eqnarray}
\psi_j''(k_j) +  2d_j^2 \left(\frac{E_j}{\hbar \omega_j} +   \frac{d_j^2}{\mu_j^2} (\cos(k_j \mu_j)-1)
 \right) \psi_j(k_j)=0\;.
\end{eqnarray}
Finally, by making the change of variables $z_j=\frac {k_j \mu_j+\pi}{2}  \in (0,\pi)$ we write
\begin{eqnarray}\label{mat}
\psi_j''(z_j)+\left[a_j-2q_j\cos(2z_j)\right]  \psi_j(z_j)=0\;,
\end{eqnarray}
where  $q_j=4\lambda_j^{-4}$, 
$a_j=\frac{8}{\lambda_j^4}  \left(\frac{\lambda_j^2E_j}{\hbar \omega_j}-1\right)$
and $\lambda_j=\mu_0/d_j$ is a dimensionless parameter.
Equation \eqref{mat} is the well-known Mathieu 
equation in its canonical form. We seek for periodic solutions of 
the Mathieu equations, since $\psi_j(0)=\psi_j(\pi)$ by construction.

In order to approximate perturbatively to the
harmonic oscillator we take $\lambda_1$ to be small, which produces the 
 the following asymptotic expansion for the first oscillator
\begin{eqnarray}
a_{1, n}(\lambda_1) &=& -8\lambda_1^{-4}+4(2 n+1)\lambda_1^{-2}-\frac 
14(2n^2+2 n+1) \nonumber \\ && +\mathcal O\left(\lambda_1^2\right) \;,
\end{eqnarray}
 for $n=0,\:1,\:2,\ldots$. Considering the asymptotic expansion for $a_1$
  we can write
\begin{eqnarray}
E_{1,{n}}=  \frac{\hbar \omega_1}{2}\left((2n+1)-\frac 1 
{16} (2n^2+2n+1)\lambda_1^{2} \right)+ \mathcal O(\lambda_1^{4}) \;. \label{E1} 
\end{eqnarray}
For the second oscillator we have two alternatives: the first one is to consider 
 $\mu_0 \ll d_2$ which analogously produces 
\begin{eqnarray}
E_{2,{m}}=  \frac{\hbar \omega_2}{2}\left((2m+1)-\frac 1 
{16} (2m^2+2m+1)\lambda_2^{2} \right)+ \mathcal O(\lambda_2^{4}) \;. \label{E2} 
\end{eqnarray}
The second alternative is to consider a large $\lambda_2$ leading to the expansion
\begin{eqnarray}
E_{2,{m}}&=&\frac{\hbar \omega_2}{\lambda_2^2} + \hbar \omega_2   
\frac{\lambda_2^{2} }{8}m^2+\mathcal O(\lambda_2^{-6}) \;. 
\end{eqnarray}
\begin{figure*}
\centering
\includegraphics[width=0.40 \textwidth]{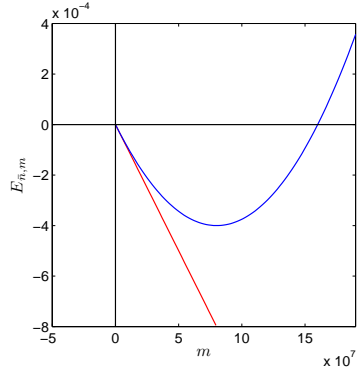}
\caption{\label{Fig1} Comparison of the P-U energies as a function of $m$ evaluated at $\bar n$ in
the Schr\"odinger and polymeric representation.}
\end{figure*}
\begin{figure*}
\centering
\includegraphics[width=0.45 \textwidth]{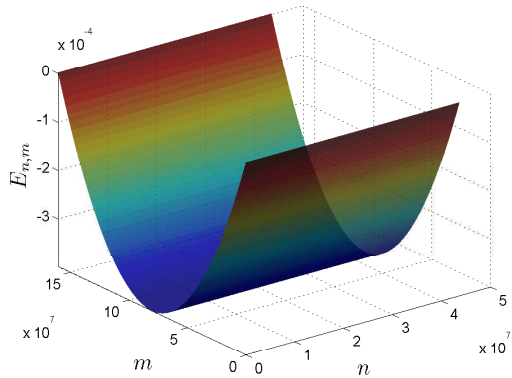}
\caption{\label{Fig2} The polymeric P-U energy $E_{n,m}$ as a function of $n$ and $m$.}
\end{figure*}
With exception of the ground state we have that 
the negative energy of the system increases without limit.
This case may be interesting to analyze from a field point of view, since in this case the propagators 
turn to be suppressed by the high scale~\cite{propagators}.
For the rigid rotator case in which $\lambda_1$ is large as well as $\lambda_2$, achieved for example taking 
large values of the original mass parameter $m$, one can see that the 
energy spectrum goes as $E_{n,m}\approx \frac{\hbar}{8} (\lambda_1^2n^2-  \lambda_2^2m^2)$ which can be more negative 
with respect to the Schr\"odinger representation for the same occupation numbers $(n,m)$. In this case there is no improvement for stability.
Let us focus on
the case $\lambda_1\ll 1$ and $ \lambda_2\ll1$ for which the total energy $E_{n,m}$ of the system
 can be written as
\begin{eqnarray}\label{En}
E_{n,m} &=&  \frac{\hbar \omega_1}{2}\left((2n+1)-\frac 1 
{16}(2n^2+2n+1)\lambda_1^2  \right)\nonumber \\ &-&\frac{\hbar \omega_2}{2}\left((2m+1)-\frac 1 
{16}(2m^2+2m+1)\lambda_2^2 \right).
\end{eqnarray}
Analogously to the Schr\"odinger quantization the 
high energy oscillator seems to lead to the instability,
however we 
will show below that in certain regions the Hamiltonian can be defined
well bounded from below. 
It is important to emphasize that in the absence of operators connecting the two Hilbert spaces
the negative energy is not to serious and the problem of instability appears upon introducing the interactions.

Let us consider the constraints imposed on the number of normal 
particles $n$
that follows from the absence of any polymeric effect in quantum mechanics. 
From the first term in \eqref{En}, it can be seen that
the positive-oscillator corrections become 
 significant or $\mathcal  O(1)$
 at the value $\bar n=\sqrt{h_1}$.
 For example considering the vibrational modes of a carbon monoxide molecule with
mass $m = 10^{-26}$ kg and frequency $ \omega_1=10^{15} \text{s}^{-1}$,
 and estimating the polymer scale to be $\mu_1 =10^{-19}$ m, 
 we find $\bar n\approx10^7$, see~\cite{ash}. 
 In addition, in several approaches to polymer quantum mechanics a cutoff in the energy eigenvalues 
 has been justified for the harmonic oscillator in order to implement a consistent renormalization~\cite{corichi}.
 This upper limit is key to
 define our effective region, since now,  in contrast to what happens in the usual P-U model
 for higher values of the occupation number 
 $m$ and for fixed $n=\bar n$ the polymeric energy is bounded from below, see Fig.~\ref{Fig1}.
 In the general setting where both occupation numbers $n$ and $m$ vary freely the total energy of the system
 is well bounded from below, avoiding possible stability problems, as shown  in Fig.~\ref{Fig2}.
\section{Conclusions}
In this work we have analyzed the stability of theories containing 
higher-order time derivatives within the framework of polymeric quantization.
For this we have focused on the well-known Pais-Uhlenbeck model with fourth-order time derivatives
in the Lagrangian.  
Using a canonical transformation we have cast the theory
into a sum of two decoupled harmonic oscillators, one with 
positive energy and the other with negative energy. 
The negative-energy oscillator is responsible for the instabilities that arise in the presence of interactions.

We have shown that the discrete nature of
the polymer Hilbert space introduces
 corrections in the energy spectrum of the P-U model
which allows to define a region of positive defined Hamiltonian. 
For this we have set a cutoff
from observational constraints from quantum mechanics
due to the absence of any polymeric 
effect and further motivated by a consistent 
renormalization program.
We have 
established an effective region defined by small values of the parameters $\lambda_1$
and $\lambda_2$ at which 
the theory has a well bounded Hamiltonian. However, for the case with 
 large $\lambda_1$ and $\lambda_2$ , we have found that the instability shows up for very low occupation numbers
with no improvement with respect to the usual Schr\"odinger quantization.
We leave  
for future investigations the inclusion of interactions and the case of large $\lambda_2$
in the context of quantum field theory.
\section*{Acknowledgments}
We want to thank H. A. Morales-Tecotl and 
T. Pawlowski
for valuable comments on this work.
P.C acknowledges support from
 Centre for Biotechnology and Bioengineering under PIA-Conicyt
 Grant No. FB0001. C.M.R. acknowledges
 support from Grant Fondecyt No. 1140781, DIUBB No. 141709 4/R and the group of \emph{Fisica de Altas Energias} of the 
Universidad del Bio-Bio.

\end{document}